\documentclass[11pt]{article}
\usepackage[margin=1.0in]{geometry}
\usepackage[printwatermark]{xwatermark}

\usepackage{verbatim}
\usepackage{amsmath}
\usepackage{amssymb}
\usepackage{multirow}
\usepackage{apacite}

\usepackage{xcolor}
\usepackage{tabu}
\usepackage{float}
\usepackage{graphicx}

\usepackage[utf8]{inputenc}
\usepackage{gensymb}
\usepackage[section]{placeins}
\usepackage{natbib}
\usepackage{lineno}
\usepackage{authblk}

\usepackage{titlesec}
\titleformat*{\section}{\large\bfseries}
\titleformat*{\subsection}{\normalsize\bfseries}

\usepackage{setspace}
\usepackage{etoolbox}
\preto\longtable{\par\singlespacing}

\usepackage{hyperref}
\hypersetup{
	colorlinks = true,
	linkcolor = {red},
	citecolor = {blue},
	linkbordercolor = {white},
	citebordercolor = {purple}
}

\title{Influence of the projectile geometry on the momentum transfer from a kinetic impactor and implications for the DART mission}

\author[1]{S. D. Raducan}
\author[1]{M. Jutzi}
\author[2]{T. M. Davison}
\author[3]{M. E. DeCoster} 
\author[3]{D. M. Graninger}
\author[4]{J. M. Owen} 
\author[3]{A. M. Stickle}
\author[2]{G. S. Collins}

\affil[1]{Space Research and Planetary Sciences, Physikalisches Institut, University of Bern, Switzerland;}
\affil[2]{Impacts and Astromaterials Research Centre, Department of Earth Science and Engineering, Imperial College London, United Kingdom;}
\affil[3]{The Johns Hopkins University Applied Physics Laboratory, Laurel, MD, United States.}
\affil[4]{Lawrence Livermore National Laboratory, Livermore, CA, United States.}
\date{}
\begin{document}

\maketitle

\begin{abstract}
The DART spacecraft will impact Didymos's secondary, Dimorphos, at the end of 2022 and cause a change in the orbital period of the secondary. For simplicity, most previous numerical simulations of the impact used a spherical projectile geometry to model the DART spacecraft. To investigate the effects of alternative, simple projectile geometries on the DART impact outcome we used the iSALE shock physics code in two and thee-dimensions to model vertical impacts of projectiles with a mass and speed equivalent to the nominal DART impact, into porous basalt targets. 
We found that the simple projectile geometries investigated here have minimal effects on the crater morphology and momentum enhancement. Projectile geometries modelled in two-dimensions that have similar surface areas at the point of impact, affect the crater radius and the crater volume by less than 5\%. In the case of a more extreme projectile geometry (i.e., a rod, modelled in three-dimensions), the crater was elliptical and 50\% shallower compared to the crater produced by a spherical projectile of the same momentum. The momentum enhancement factor in these test cases, commonly referred to as $\beta$, was within 7\% for the 2D simulations and within 10\% for the 3D simulations, of the value obtained for a uniform spherical projectile. 

The most prominent effects of projectile geometry are seen in the ejection velocity as a function of launch position and ejection angle of the fast ejecta that resides in the so-called `coupling zone'. These results will inform the LICIACube ejecta cone analysis.  
\end{abstract}
\clearpage

\section{Introduction}
NASA’s Double Asteroid Redirection Test (DART) will impact the secondary of the 65803 Didymos binary asteroid system, Dimorphos, at the end of 2022 \citep{Cheng2018}. The impact will cause a measurable change in the orbital period of Dimorphos around the primary. Several years after the DART impact, ESA's Hera mission will arrive at the system and will characterise the binary system in detail, particularly Dimorphos and the crater produced by DART on its surface \citep{Michel2018}.

When deflecting an asteroid by the means of a kinetic impactor, the amount by which the asteroid is deflected is determined by the target properties (e.g., cohesion, porosity, internal friction) and impact conditions (e.g., impact speed, projectile characteristics, impact angle). In the case of the DART impact, the impact conditions are mostly known; the mission design dictates the projectile shape and mass, the impact speed and to some extent the impact angle. However, most of the target properties will remain unknown until after the DART impact. For this reason, recent numerical simulations of the DART impact have focused on determining the sensitivity of the ejecta momentum transfer and the asteroid deflection efficiency to surface material properties and asteroid structure \citep[e.g.,][]{Jutzi2014, Stickle2015, Syal2016, Raducan2019, Raducan2020}. For simplicity and computational expediency, these studies have used either a uniform solid or porous aluminium sphere as the impactor. However, the DART spacecraft's structure and mass distribution is more complex and the effects caused by the spacecraft geometry on the crater size and morphology and on the momentum enhancement are not yet known.

Previous laboratory studies suggest that impactor geometry can have an effect on the ejecta generated from a high velocity impact. For example, \cite{Hermalyn2012} conducted laboratory experiments of solid and hollow aluminium projectiles impacting sand and pumice targets at $\approx$~2.5\,km/s and showed that the two projectiles types produce ejecta with different velocity and ejection angle distributions as a function of launch position. The solid projectile produced ejecta that had a significantly steeper ejection angles compared to the ejecta produced by the hollow projectiles.  

In high velocity impacts on an asteroid the change in momentum of the asteroid, $\Delta P$, can be amplified by the momentum of crater ejecta that exceeds the escape velocity, which is often expressed in terms of the parameter $\beta = \Delta P/mU$, where $mU$ is the impactor momentum \citep{Housen2012}. Therefore, in the case of a kinetic impactor, a lower ejection angle would imply a lower vertical momentum component, and hence a lower $\beta$ value \citep{Raducan2021c}.

Here we used numerical impact simulations in two and three dimensions to quantify the effects of simple projectile geometries on the crater size and morphology, and on the ejecta momentum transfer. One of the aims of this study is to understand the role of simplified projectile geometries on the effects of a kinetic impact. Such studies are important to determine whether simple projectile structures are appropriate when modelling a kinetic impactor, which is the subject of ongoing investigations.

\section{Numerical model}\label{numerical_model}
We used the iSALE shock physics code in two (-2D) and three dimensions (-3D) \citep{Wunnemann2006, Elbeshausen2011} to model DART-like impacts on asteroid surfaces, considering different projectile geometries. iSALE is a multimaterial, multirheology extension of the SALE hydrocode \citep{Amsden1980}, specifically developed for simulating impact processes and similar to the older SALEB hydrocode \citep{Ivanov1997, Ivanov_Artemieva2002}. iSALE-3D \citep{Elbeshausen2009, Elbeshausen2011} uses a 3D solution algorithm very similar to the SALE-2D solver, as described by \cite{Hirt1974}. The development history of iSALE-3D is described in \cite{Elbeshausen2009}. Both codes share the same material modelling routines, including strength models suitable for impacts into geologic targets \citep{Collins2004} and a porosity compaction model \citep{Wunnemann2006}. 

iSALE-2D has been extensively validated against laboratory impact experiments \citep{Wunnemann2016}, and benchmarked against other hydrocodes \citep{Pierazzo2008, Davison2011, Stickle2020b} for simulating the crater size and morphology. Moreover, the ejection velocities and angles produced by vertical impacts have been shown to be in good agreement with data from laboratory impacts into sand \citep{Luther2018, Raducan2019} and regolith simulant \citep{Raducan2021}. The crater sizes produced in iSALE-3D impact simulations into aluminium targets also showed good agreement with laboratory data \citep{Davison2011}.

This numerical study is divided into two parts that aim to quantify the effects of projectile geometry on crater size, ejecta mass-velocity distribution and momentum enhancement: iSALE-2D simulations, in two dimensions, with an axially symmetric geometry; and iSALE-3D simulations, in three dimensions,  which employ Cartesian coordinates ($x-y-z$). For these simulations, the computational domain was modelled as a half-space, with a symmetry axis along the horizontal component of the projectile velocity, in the $x-z$ plane.

The projectiles used in this study are made of non-porous aluminum with a given shape and were modelled using the Tillotson equation of state (EoS) \citep{Tillotson1962} and the Johnson-Cook strength model \citep{Johnson1983}. The projectile input parameters are summarised in Table~\ref{table:model_parameters}.  In all simulations, the impact speed was kept constant, at 6.5 km/s, and the gravitational acceleration, at 5$\times10^{-5}$ m/s$^2$.

\subsection{Projectile geometry in two-dimensions (2D)}

\begin{figure}[!h]
	\centering
	\includegraphics[width=0.65\linewidth]{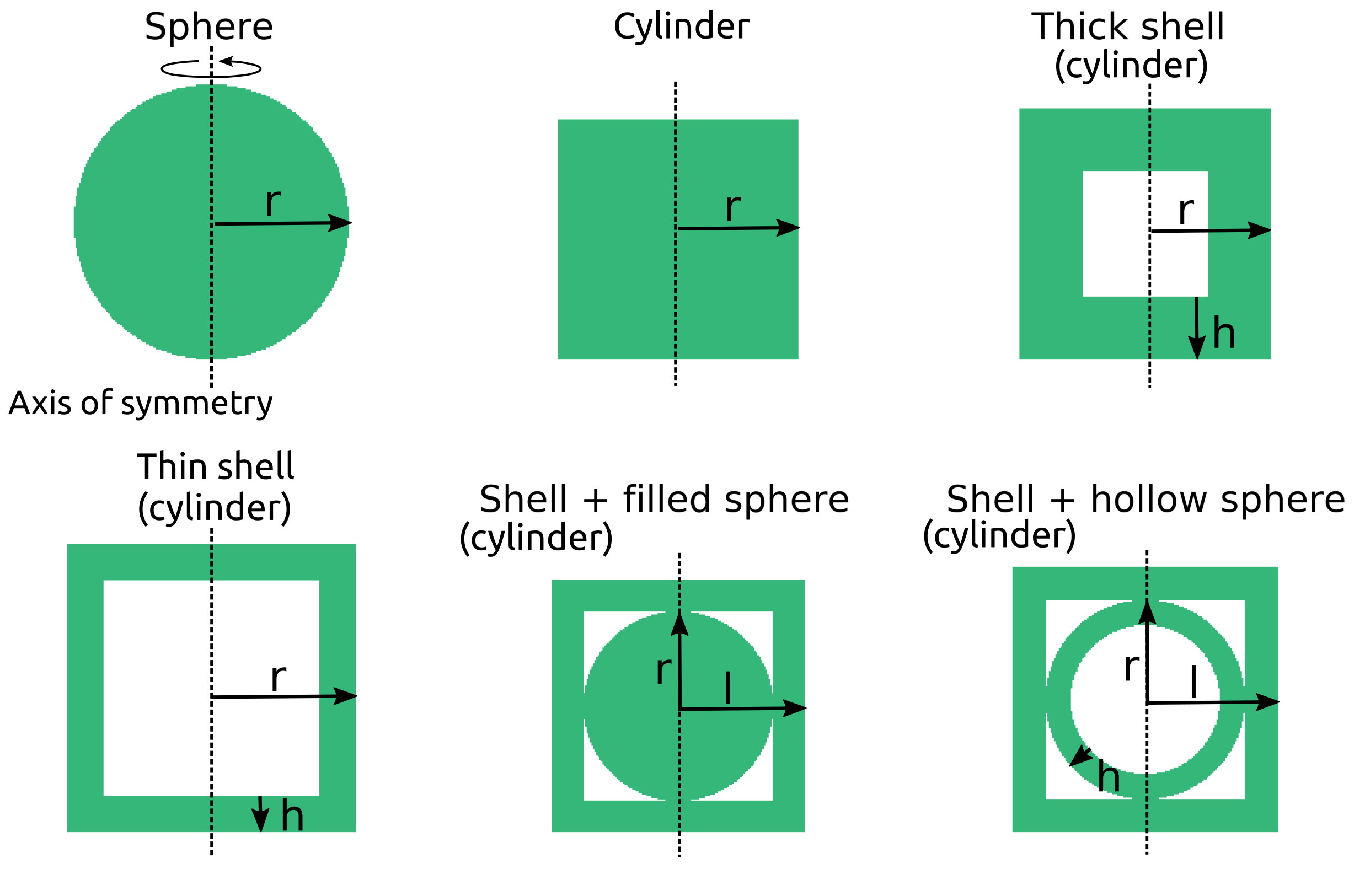}
	\caption{Schematic representation of the projectile structures considered in the iSALE-2D study.}
	\label{fig:imp}
\end{figure}

Two-dimensional simulations with simple projectile structures provide a method for building up intuition about the spacecraft geometry. Because 2D simulations are less computationally expensive than the 3D equivalents, they allow us to test a larger number of initial conditions and at higher initial spatial resolution. In order to determine the final crater size and morphology, yet capture the subtle differences in the ejecta mass-velocity distribution, all 2D simulations presented here used regridding \citep{Raducan2019}. A variety of simple shapes were chosen to represent the DART spacecraft bus (Fig~\ref{fig:imp}): sphere, cylinder, cylinder with a thick shell, cylinder with a thin shell, cylinder with a thin shell and an enclosed filled sphere and a cylinder with a thin shell and with an enclosed hollow sphere. We note that the projectile structures considered in this study had a similar contact surface area at impact. All the structures considered here were made of solid aluminium and had constant mass, $m$ = 650\,kg. The projectile geometries used in this study are shown in Table~\ref{table:impactor_2d}. 
\vspace*{-0.4cm}
\begin{table}[ht!]
	\caption{Table of impactor input parameters from iSALE-2D simulations (cppr = cells per projectile radius)}.
	\begin{tabular}{l@{\hskip 0.25in}c@{\hskip 0.25in}c@{\hskip 0.25in}c@{\hskip 0.25in}}
		\hline
        \textbf{Shape} &\textbf{Dimensions}  & \textbf{Initial resolution}  & \textbf{Mass} \\
		\hline
	   Sphere                 & $r$ = 38.6 cm (80 cppr) &   4.82 mm/cppr   & 650 kg \\
	   \hline
	   Cylinder                    & $r$ = 33.7 cm (80 cppr) &   4.21 mm/cppr   & 650 kg \\
	   \hline
	   \multirow{2}{*}{Thick shell (cylinder)} & $r$ = 35.2 cm (80 cppr) & \multirow{2}{*}{4.41 mm/cppr }&  \multirow{2}{*}{650 kg} \\
	                                & $h$ = 17.6 cm (40 cppr) &                               &                          \\         
	                               
	   \hline                            
	   \multirow{2}{*}{Thin shell (cylinder)} & $r$ = 40.5 cm  (80 cppr) & \multirow{2}{*}{5.06 mm/cppr }&  \multirow{2}{*}{650 kg} \\
	                               & $h$ = 10.1 cm  (20 cppr) &                               &                          \\         
	                               
	   \hline     
	   \multirow{2}{*}{Shell (cylinder) + filled sphere} & $l$ = 35.4 cm (80 cppr) & \multirow{2}{*}{4.43 mm/cppr }&  \multirow{2}{*}{650 kg} \\
	                                          & $r$ = 17.7 cm (40 cppr) &                               &                          \\      
	   
	   \hline
       \multirow{3}{*}{Shell (cylinder) + hollow sphere} & $l$ = 37.3 cm  (80 cppr) & \multirow{3}{*}{4.66 mm/cppr }&  \multirow{3}{*}{650 kg} \\
	                               & $r$ = 28.0 cm (60 cppr) &                               &                          \\      
	                               & $h$ = 7.0 cm  (15 cppr) &                               &                          \\      
	   
	   \hline
	\end{tabular}
	\label{table:impactor_2d} 
\end{table}

\subsection{Projectile geometry in three-dimensions (3D)}

The second study considered two projectile geometries modelled in three-dimensions (3D): a $\approx$~572\,kg solid aluminium sphere and a $\approx$~572\,kg horizontal solid aluminium rod (impacting vertically). Because of the low spatial resolution (5 cppr) employed to model the 3D impact scenarios here, the projectile is less well resolved than the projectiles in the 2D study. Therefore,  the projectile mass was slightly lower than the theoretical, assumed $\approx$~600\,kg mass (see Table~\ref{impactor_3d}). These two projectile geometries are extreme cases intended to represent two different idealised spacecraft geometries (i.e., the rod is a simplification of a spacecraft with solar panels), and are chosen to maximise possible differences in crater size and morphology and ejecta mass-velocity distribution. The iSALE-3D simulations presented here had a much coarser spatial resolution compared to the iSALE-2D study. 

\begin{table}[ht!]
	\caption{Table of impactor input parameters from iSALE-3D simulations (cppr = cells per projectile radius)}.
	\begin{center}
	   
	\begin{tabular}{l@{\hskip 0.5in}c@{\hskip 0.5in}c@{\hskip 0.5in}c@{\hskip 0.5in}}
		\hline
        \textbf{Shape} &\textbf{Dimensions}  & \textbf{Resolution}  & \textbf{Mass} \\
		\hline
	   Sphere                 & $r$ = 37.5 cm (10 cppr) &   2.5 cm/cppr   & 572 kg \\
	   \hline
       \multirow{2}{*}{Rod} & $l$ = 4.42\,m  (200 cppr) & \multirow{2}{*}{2.21 cm/cppr }&  \multirow{2}{*}{575 kg} \\
	                               & $h$ = 11.05\,cm (5 cppr) &                               &                          \\     
	   \hline
	\end{tabular}
	\end{center}
	\label{impactor_3d} 
\end{table}

\subsection{Target material model}

For consistency, in all impact scenarios considered here, the target was modelled using the same material parameters and only varying the target strength. Observational studies suggest that Didymos is an S-type silicaceous asteroid system \citep[e.g.,][]{DeLeon2006}, however here the target asteroid material was considered to be made of weak porous basalt, considered to be a good approximation of most asteroids. Similar analogue material has been used in previous studies of asteroid impacts \citep[e.g.,][]{Durda2007, Benavidez2012, Jutzi2014, Stickle2015, Jutzi2019}. Therefore, the target was modelled using the Tillotson equation of state for basalt \citep{Tillotson1962, Benz1999}, which is coupled to the $\epsilon-\alpha$ porosity model \citep{Wunnemann2006, Collins2011}, and a pressure-dependent strength model \citep{Collins2004}. The porosity was kept constant at 20\% porosity, which is the current best estimate of the primary's porosity \citep{Naidu2020}. To study the influence of the projectile geometry on the cratering process, given different cratering efficiencies, here we modelled targets with varying cohesive strength. 

The cohesive strength of the target material was modelled using iSALE's ROCK model \citep{Collins2004}, a complex strength model in which a pressure-dependent strength is reduced as strain accumulates, and the yield strength is defined as:
\begin{equation}
    Y = Y_d D +Y_i\left(1-D \right)
\end{equation}
where $D$ is a scalar measure of damage, which is itself a function of accumulated plastic strain. The strength of the damaged material, $Y_d$ is defined by a Drucker-Prager relationship:
\begin{equation}
    Y_d = \min\left(Y_{d0}+f_dP, Y_{dm} \right),
\end{equation}
and the intact material strength, $Y_i$, is defined by the Lundborg relationship:
\begin{equation}
    Y_i = Y_{i0} + \frac{f_iP}{1+f_iP/(Y_{im}-Y_{i0})},
\end{equation}
where $f_i$ is the coefficient of internal friction for intact material and $f_d$ is the coefficient of internal friction for damaged material. The assumption used in the ROCK strength model is that the strength of the intact and damaged material are the same above the brittle-ductile transition. For this reason, here the limiting strength at high pressure for intact material and the limiting strength for damaged material at high pressure are chosen to have the same value: $Y_{im} = Y_{dm} = Y_{inf}$. The plastic strain was related to the damage, $D$, using the damage model described by \cite{Ivanov2010}.

In our iSALE-2D impact simulations, we considered two targets that were intact before the impact ($D$ = 0) and a weaker target, that was fully-damaged before the impact ($D = 1$). In the intact target scenarios, the intact material strength, $Y_{i0}$, was varied between 1 MPa and 100 MPa, and the cohesive strength of the damaged material, $Y_{d0}$, was varied between 100 kPa and 10 MPa, respectively. The fully-damaged target had $Y_{d0}$ = 0.1 kPa. In our iSALE-3D impact simulations, the target was intact before the impact and had $Y_{i0}$ = 90 MPa and $Y_{d0}$ = 1 MPa. The target input parameters are summarised in Table~\ref{table:model_parameters}. 

As described in \cite{Raducan2019}, it is the post-shock, damaged strength of the asteroid surface, rather than the intact strength of the material that exerts most influence on the crater ejecta behaviour. Therefore, to simplify the notation throughout this paper, we refer to the different target cohesion configurations described above by their damaged strength, as follows: $Y_{d0}$ = 1 kPa, 0.1 MPa and 10 MPa for the iSALE-2D simulations and $Y_{d0}$ = 1 MPa for the iSALE-3D simulations. 

To be able to simulate a large number of projectile geometries with high spatial resolution, here we consider relatively strong targets, that have a low cratering efficiency. Even though the cohesions used here are higher than the values recently measured on asteroid surfaces (e.g., the recently visited asteroids Ryugu and Bennu have surfaces with very low cohesions \citep{Arakawa2020}), these targets are still suitable to inform possible projectile geometry effects on the cratering process. While there will be an interplay between projectile geometry effects and target property effects, projectile geometry effects will be most evident in impacts with low cratering efficiency, where the final crater is not much larger than the projectile size. Conversely, projectile geometry effects are expected to be greatly diminished in higher cratering efficiency impacts where the crater grows to many times larger than the projectile. Projectile geometry effects will also be most evident in the fastest ejecta, produced at very early times close to the contact zone when target strength effects are negligible because the driving stresses are so large. Hence, projectile geometry effects on fast ejecta are expected to be independent of target strength. As $\beta$ is dominated by the slowest ejecta, projectile effects on deflection efficiency are also expected to diminish with decreasing target strength. Here we focus on relatively high cohesion targets to maximise the potential influence of projectile geometry. Our results should therefore provide an upper bound on the potential influence of projectile geometry effects.  

\begin{table}[h]
    \footnotesize
	\caption{Material model parameters for DART-like impact. For all simulated materials we used the thermal parameters from \cite{Ivanov2010}. }
	\begin{tabular}{l@{\hskip 0.1in}l@{\hskip 0.2in}l@{\hskip 0.2in}l}
    Description        & Impactor & iSALE-2D target & iSALE-3D target  \\
    \hline
    Material             & Aluminium & Basalt & Basalt\\
	Equation of state        & Tillotson$^{a}$ & Tillotson$^b$  & Tillotson$^b$\\
	Strength model           & Johnson-Cook & ROCK$^{c}$ & ROCK$^{c}$  \\
	Poisson ratio,     $\nu$ & 0.33      & 0.25     & 0.25 \\
	\hline
	ROCK strength parameters$^{c}$ \\
    Intact strength at zero pressure, Y$_{i0}$ (MPa) & --  & -/1/100  & 90 \\
	Internal friction coefficient (intact), $f_i$    & --  & 1.2 & 1.2 \\
	
	Damage strength at zero pressure, Y$_{d0}$ (kPa)& --  & 1/100/10$^4$  & 100 \\
	Strength at infinite pressure, Y$_{\inf}$ (GPa) & --  & 3.5 & 3.5 \\
	Internal friction coefficient (damaged), $f_d$  & --  & 0.6 & 0.6 \\
	\hline
	Johnson-Cook strength parameters$^{d}$ \\
    Strain coefficient A (MPa), $A$      & 244  & -- & -- \\
	Strain coefficient B (MPa), $B$      & 488  & -- & -- \\
	Strain exponent, $n$                 & 0.50 & -- & -- \\
	Strain rate coefficient, $C$         & 0.02 & -- & -- \\
	Thermal softening, $m$               & 1.7 & -- & -- \\
	\hline
	Porosity model parameters ($\epsilon-\alpha$)$^e$           \\ 
	Initial porosity, $\phi_0$             & - & 20\% & 20\% \\
	Initial distension, $\alpha_0$         & - & 1.25 & 1.25 \\
	Distension at transition to power-law, $\alpha_x$ & - & 1.00 & 1.00 \\
	Elastic volumetric strain threshold, $\epsilon_{e0}$ & - & -1.88$\times 10^{-4}$ & -1.88$\times 10^{-4}$ \\
	Exponential compaction rate, $\kappa$        & - & 0.90 & 0.90 \\
	\hline
    \multicolumn{4}{l}{
    $^a$\cite{Tillotson1962};
    $^b$\cite{Benz1999};
    $^c$\cite{Collins2004};
    $^d$\cite{Pierazzo2008};}\\
    {$^e$\cite{Wunnemann2006}.}
	\end{tabular}
	\label{table:model_parameters}
\end{table}

\subsection{Ejecta measurement and resolution tests}
In iSALE, tracer particles are placed in the high resolution domain and their mass, velocity and radial distance are measured at a fixed altitude above the surface, here set to one impactor diameter, to exclude excavated material forming in the crater rim that never detaches from the target \citep{Luther2018, Raducan2019}. To quantify the effects of spatial resolution when resolving the impactor and the resulting ejecta, we first conducted a series of resolution tests.

\begin{figure}[!h]
	\centering
	\includegraphics[width=\linewidth]{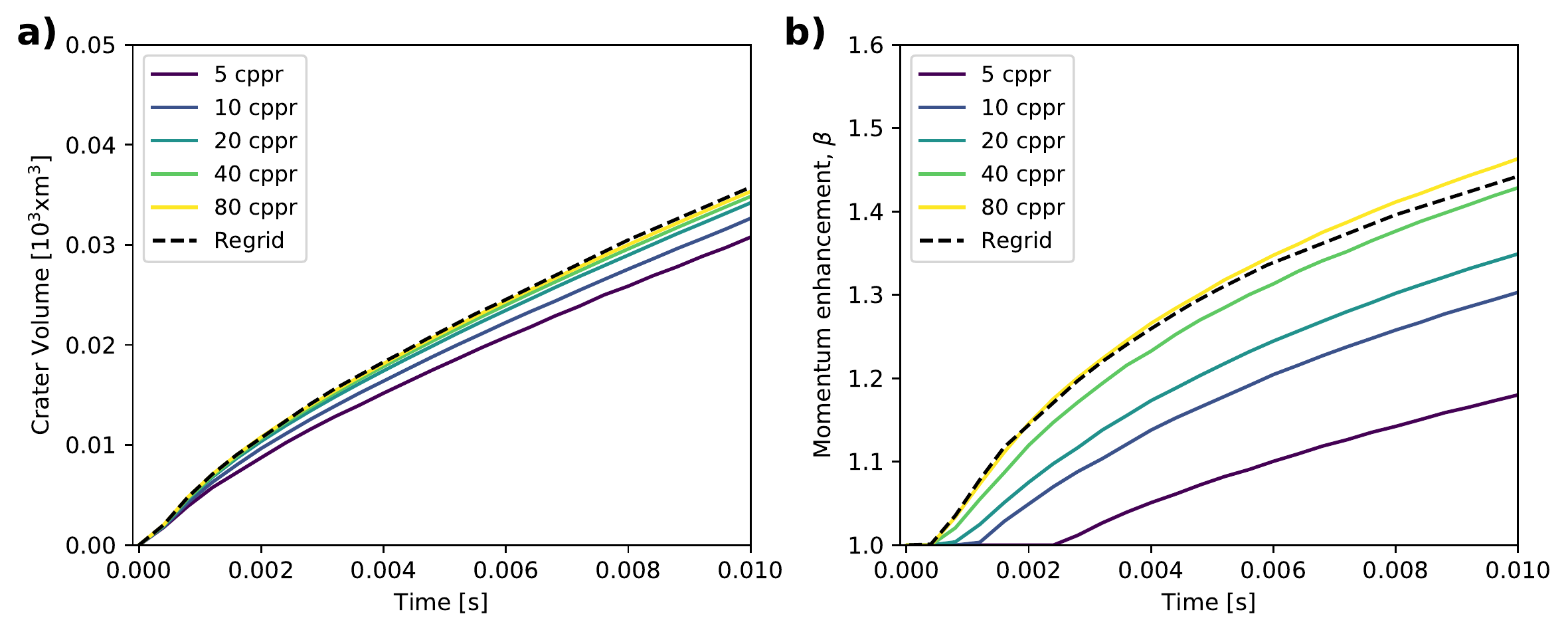}
	\caption{a) Crater volume as a function of time for a solid aluminium sphere impacting a $Y_{i0}$ = 10 MPa target, at resolutions between 80 and 5 cppr, compared to a simulation where regridding was used. b) Momentum enhancement, $\beta$ as a function of time, for varying resolution. }
	\label{fig:resolution}
\end{figure}

Figure~\ref{fig:resolution}a shows the crater volume as a function of time from iSALE-2D for a solid aluminium sphere impacting a basalt target (20\% porous and $Y_{i0}$ = 10 MPa), resolved with 5, 10, 20, 40 and 80 cells per projectile radius (cppr). At 0.01\,s after the impact, the simulation with the lower, 5 cppr, spatial resolution underpredicts the crater volume by about 12\%, compared to a simulation with 80 cppr resolution. When comparing the momentum enhancement from these simulations (Fig.~\ref{fig:resolution}b), the difference between the 5 cppr run and the 80 cppr run was almost 20\%. As discussed in \cite{Raducan2019}, when measuring the ejecta, a high spatial resolution is required in order to capture the high velocity particles. Though these particles have a low mass, they carry a significant amount of the total ejected momentum. In order to maintain the high spatial resolution needed to record the fast ejecta at the beginning of the cratering process, here we used the regridding option described in \cite{Raducan2019}. The regridding option allowed us to coarsen the simulation domain by a factor of two during the crater formation process, starting from 80 cppr at the beginning of the simulation and ending with 10 cppr by the end of the simulation. When regridding was used, the crater volume and the $\beta$ value recorded was within 2\% from the 80 cppr run (Fig.~\ref{fig:resolution}). The small difference in $\beta$ between the 80 cppr and the simulation that used regridding is due to less fast ejecta being recorded in the regridding scenario.

\subsection{Fate of projectile}

In the impact scenarios considered here, the projectiles experience extreme shock pressures and temperatures. To record the thermodynamic history of the impactor material we used Lagrangian tracer particles, which were placed in every cell across the impactor, at the beginning of the simulation (80\,cppr resolution). This allowed us to record the peak pressures and temperatures, which were then analysed in post-processing \citep{Potter2013}.

\begin{figure}[!h]
	\centering
	\includegraphics[width=\linewidth]{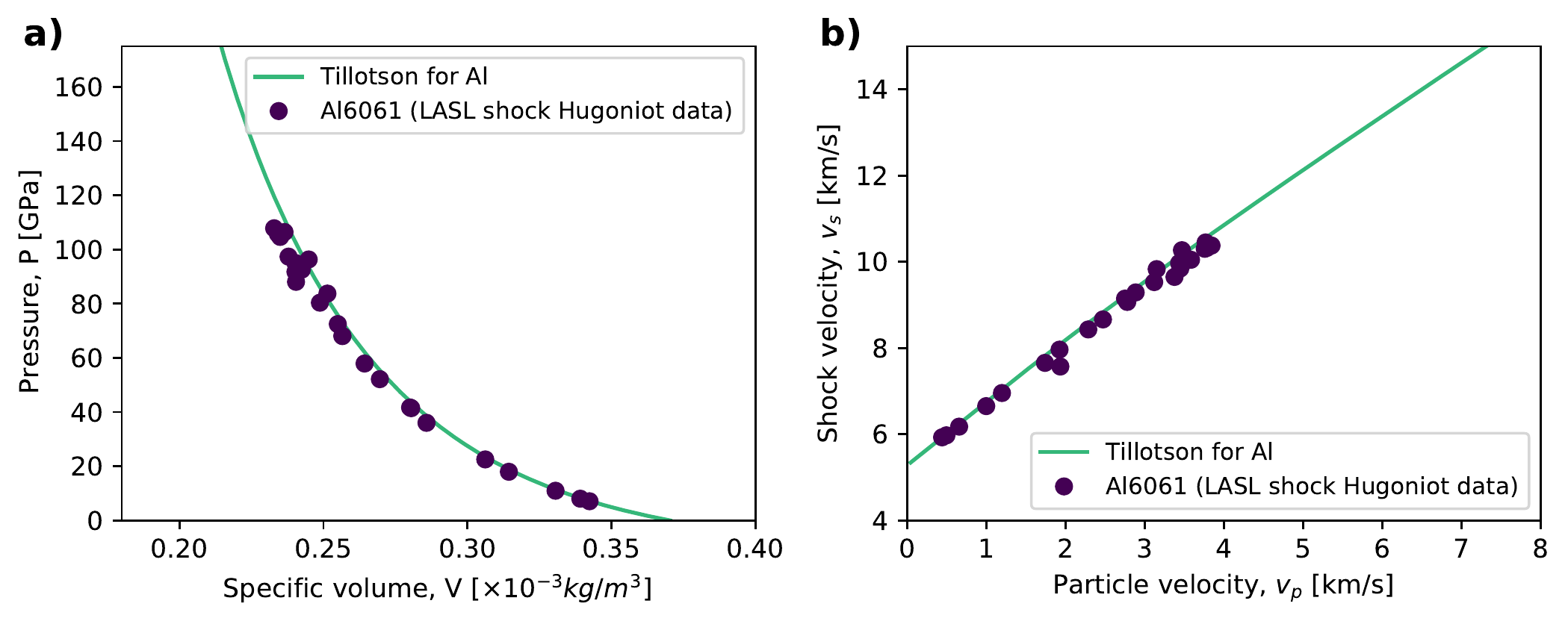}
	\caption{Tillotson Hugoniot for an aluminium projectile in the a) pressure-volume (P-V) space and in the b) particle velocity - shock velocity ($v_p - v_s$) space, compared to shock data from LASL shock Hugoniot data for Al6061 \citep{Marsh1980}. }
	\label{fig:hugoniot}
\end{figure}

The projectile material was modelled using the Tillotson EoS for aluminium \citep{Tillotson1962}, which is designed to fit experimentally derived linear shock velocity--particle velocity relations and is extrapolated to the Thomas-Fermi limit at high pressures \citep{Pierazzo1997}. While the Tillotson EoS is a relatively simple model, it is not always thermodynamically consistent and it is especially inaccurate in the partial vaporisation regime (i.e., when gas and liquid coexist) \citep{Melosh1989}. However, the Tillotson EoS for Aluminium approximates well the material behaviour at low pressures ($\lesssim$~100\,GPa) and a comparison with experimental shock data from LASL shock Hugoniot data for Al6061 \citep{Marsh1980} shows a very good agreement for these pressures (Fig.~\ref{fig:hugoniot}a), which indicates that the EoS accurately represents the target material behaviour at pressures lower than $P\lesssim$~100\,GPa. For pressures higher than $P\gtrsim$~100\,GPa, not enough experimental data is available and therefore behavior in this regime may not be well represented by the EoS. Figure ~\ref{fig:hugoniot}b shows the particle velocity as a function of shock velocity from the EoS, compared to shock Hugoniot data for AL6061 \citep{Marsh1980}.

\begin{figure}[!h]
	\centering
	\includegraphics[width=0.6\linewidth]{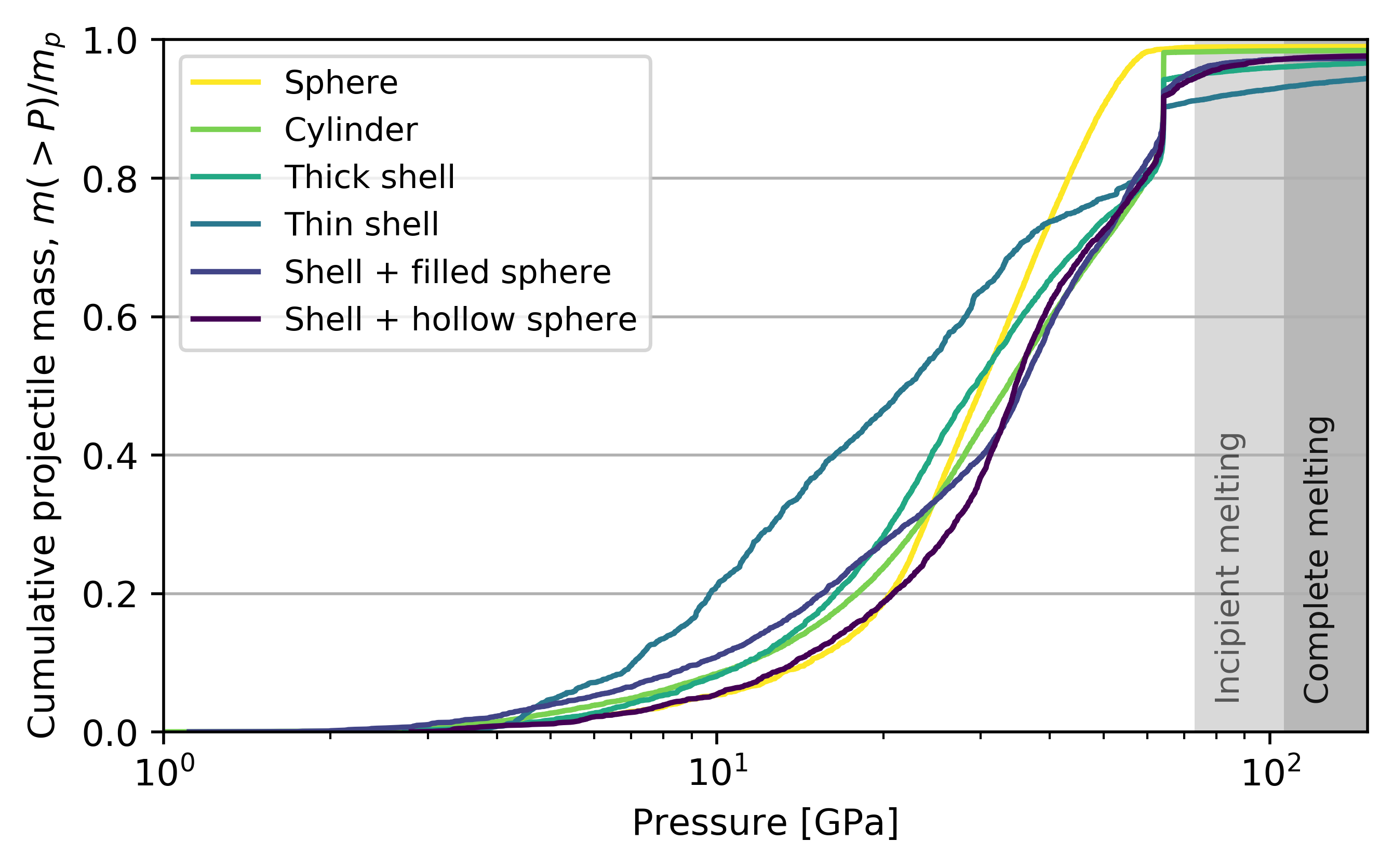}
	\caption{Cumulative mass fraction of impactor material that experiences a peak shock pressure less than a given pressure, for different projectiles impacting a 10\,MPa basalt target. The grey shaded areas represent the pressure and temperature thresholds (at 1\,bar) for aluminium incipient and complete melting \citep{Pierazzo1997}. }
	\label{fig:imp_pt}
\end{figure}

Figure~\ref{fig:imp_pt} shows the cumulative mass fraction of impactor material that experiences peak shock pressures less than a given pressure, for different shaped projectiles impacting a 0.1\,MPa basalt target, at 6.5\,km/s. All projectiles considered here experience similar peak pressures, with only small differences due to the projectile shape. In the case of the spherical projectile, all the material experiences pressures of less than $\approx$~40\,GPa, while in the other five projectile scenarios, only 80\% of the projectile material experienced peak pressures of less than 40\,GPa. 
This result indicates that there is a subtle effect of projectile geometry on the impact effects, however differences in the spatial distribution of the shock pressures are not significant enough to play a role on crater morphologies or ejected material.

The peak shock pressures required for incipient and complete melting of the projectile are given from the intersection of the Hugoniot with the aluminium isentropes on a phase diagram. The incipient and impressive melting as measured at 1\,bar are $P_{im}$ = 73\,GPa, or the complete melting pressure, $P_{cm}$ = 106\,GPa \citep{Pierazzo1997} and are plotted on Figure~\ref{fig:imp_pt} for reference. Our simulation results suggests that regardless of the projectile geometry, only a small fraction of the projectile material experiences peak shock pressures higher than the incipient melting pressure, $P_{im}$, or the complete melting pressure, $P_{cm}$. 

In the impact scenarios considered here, most of the projectile material experiences pressures of less than $P\lesssim$~100\,GPa. However, while in this region the shock compression behaviour is well represented by the Tillotson EoS, a more sophisticated EoS (e.g., ANEOS) is required for a more detailed and more accurate analysis of the thermodynamic effects. 

\section{Results from iSALE-2D simulations}

\subsection{Crater size and morphology}

\begin{figure}[!h]
	\centering
	\includegraphics[width=0.325\linewidth]{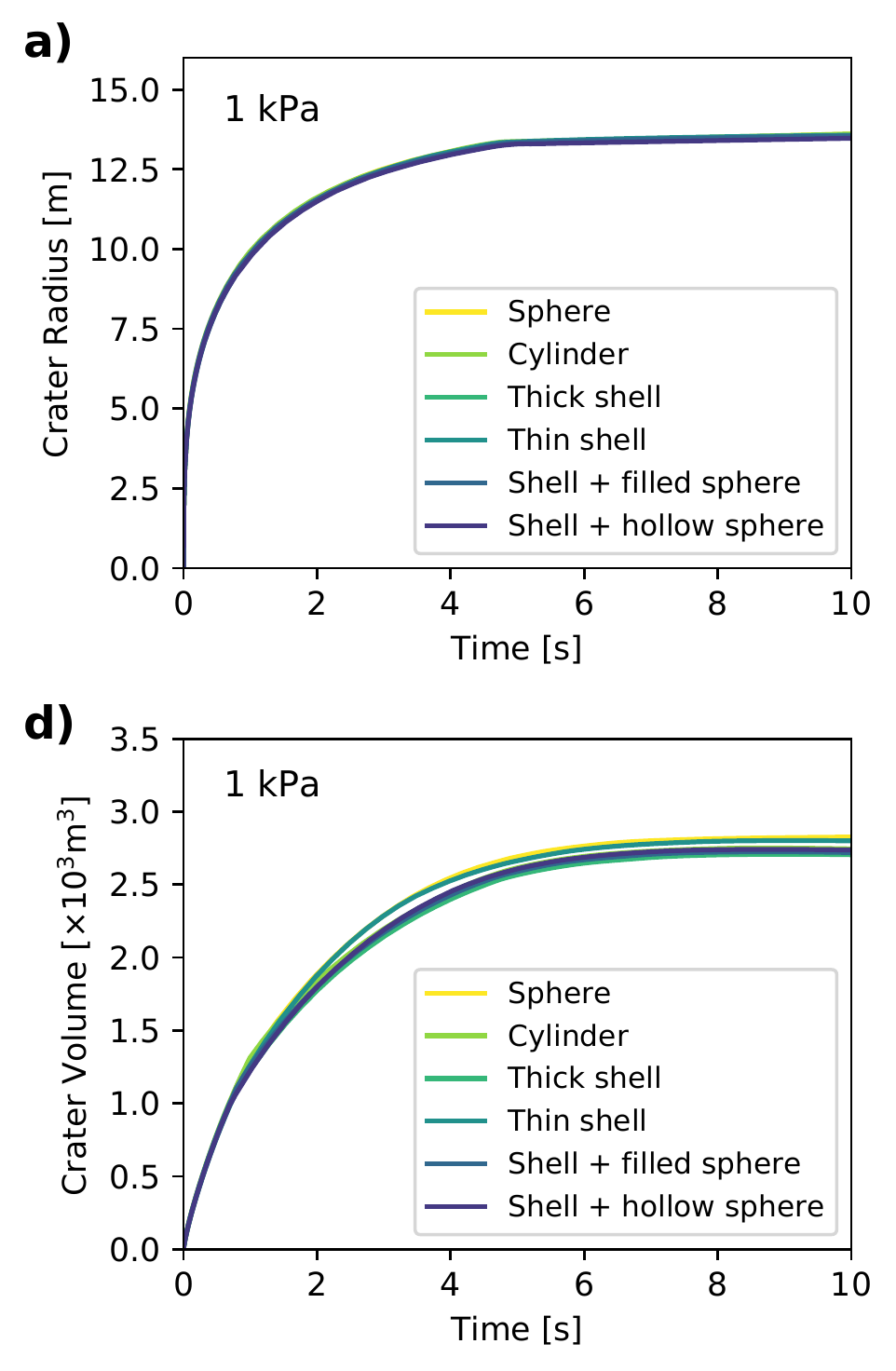}
	\includegraphics[width=0.325\linewidth]{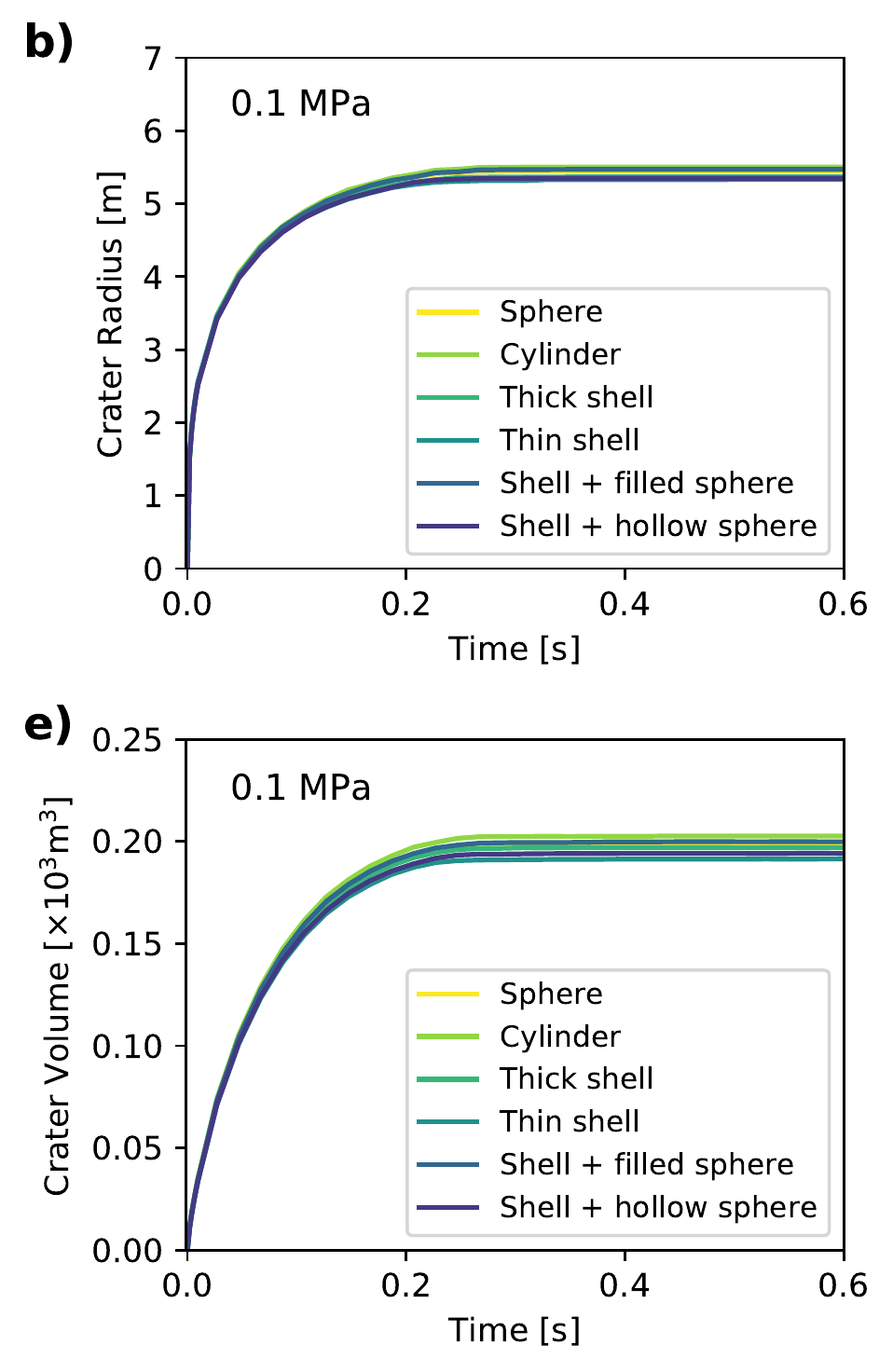}
	\includegraphics[width=0.325\linewidth]{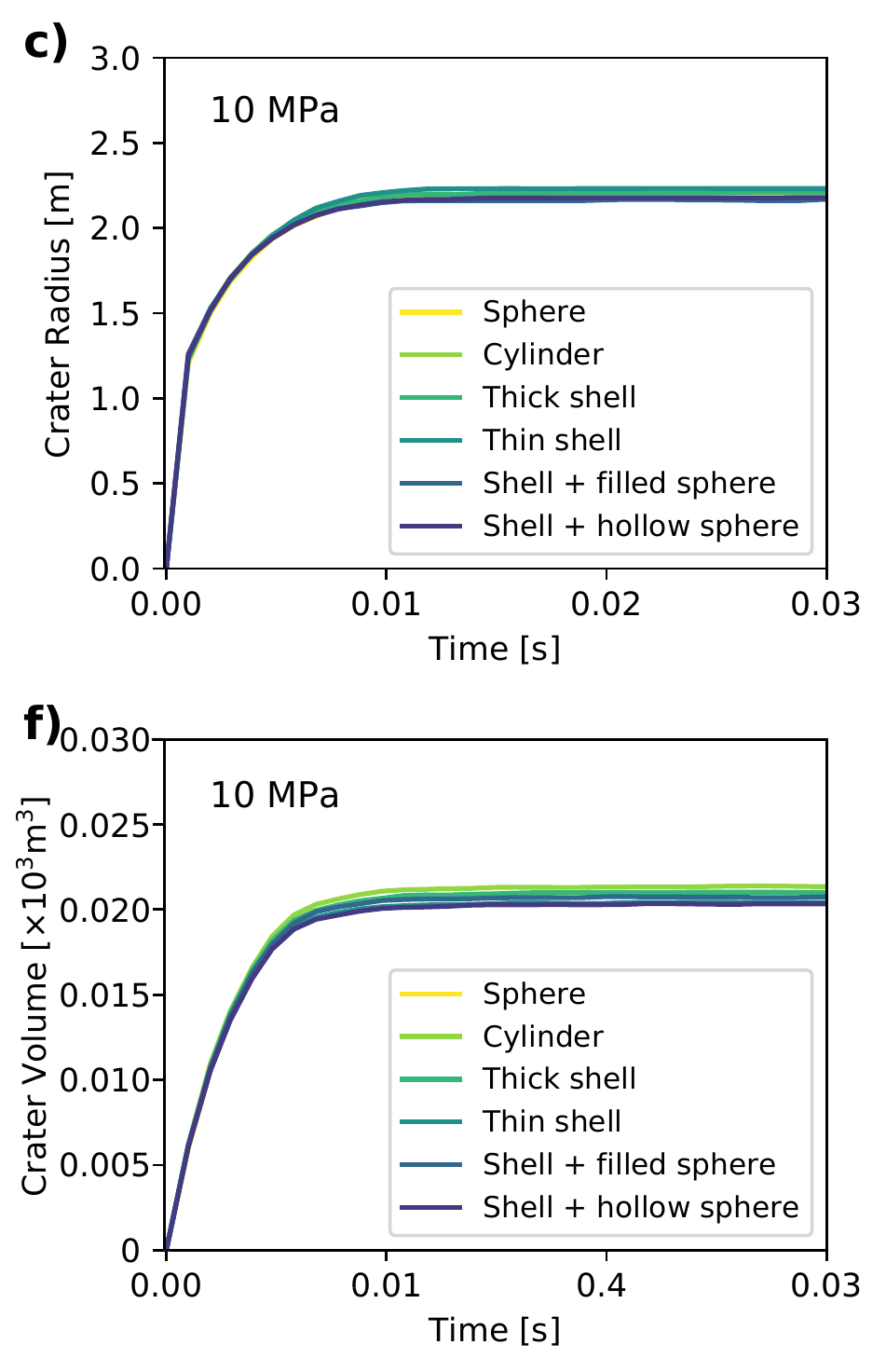}
	\caption{Crater radius and crater volume as a function of time for different shaped impactors into a 20\% porous basalt target and a cohesion of a), d) 1 kPa, b), e) 0.1 MPa and c, f) 10 MPa. The small discontinuities in a) and c) are due to time undersampling.}
	\label{fig:crater_size}
\end{figure}

Firstly, we investigated the influence of the projectile shape on the crater size. Here we modelled projectiles with six different geometries (Fig.~\ref{fig:imp}), impacting a 20\% porous basalt target. In order to determine whether the target cohesion plays a role when quantifying the effects of the projectile geometry, here we modelled impacts into targets with three different cohesions, $Y_{d0}$ = 1 kPa, 0.1 MPa and 10 MPa (see Section~\ref{numerical_model} for details on the notation), at 6.5 km/s. Figure~\ref{fig:crater_size}a, b and c show the crater radius, measured at the pre-impact level, as a function of time for the six different projectile geometries. Figure~\ref{fig:crater_size}d, e and f show the crater volume, as a function of time. The crater growth is shown for the three target cohesions, $Y_{d0}$ = 1 kPa, 0.1 MPa and 10 MPa and the crater growth time decreases with decreasing cratering efficiency, which is lower for higher target strength. For a 1\,kPa target, the crater stops growing at about 5\,s after the impact, for a 0.1\,MPa target, at about 0.3\,s after the impact and for a 10 MPa target, at about 0.003\,s after the impact. These results are consistent with crater scaling laws \citep[e.g.,][]{Housen1983, Holsapple2007}. For all three target cohesions considered here, the impacts of different projectile structures produced crater radii that were within 5\% and crater volumes that were within 3\% of each other. 

\begin{figure}[!h]
	\centering
	\includegraphics[width=\linewidth]{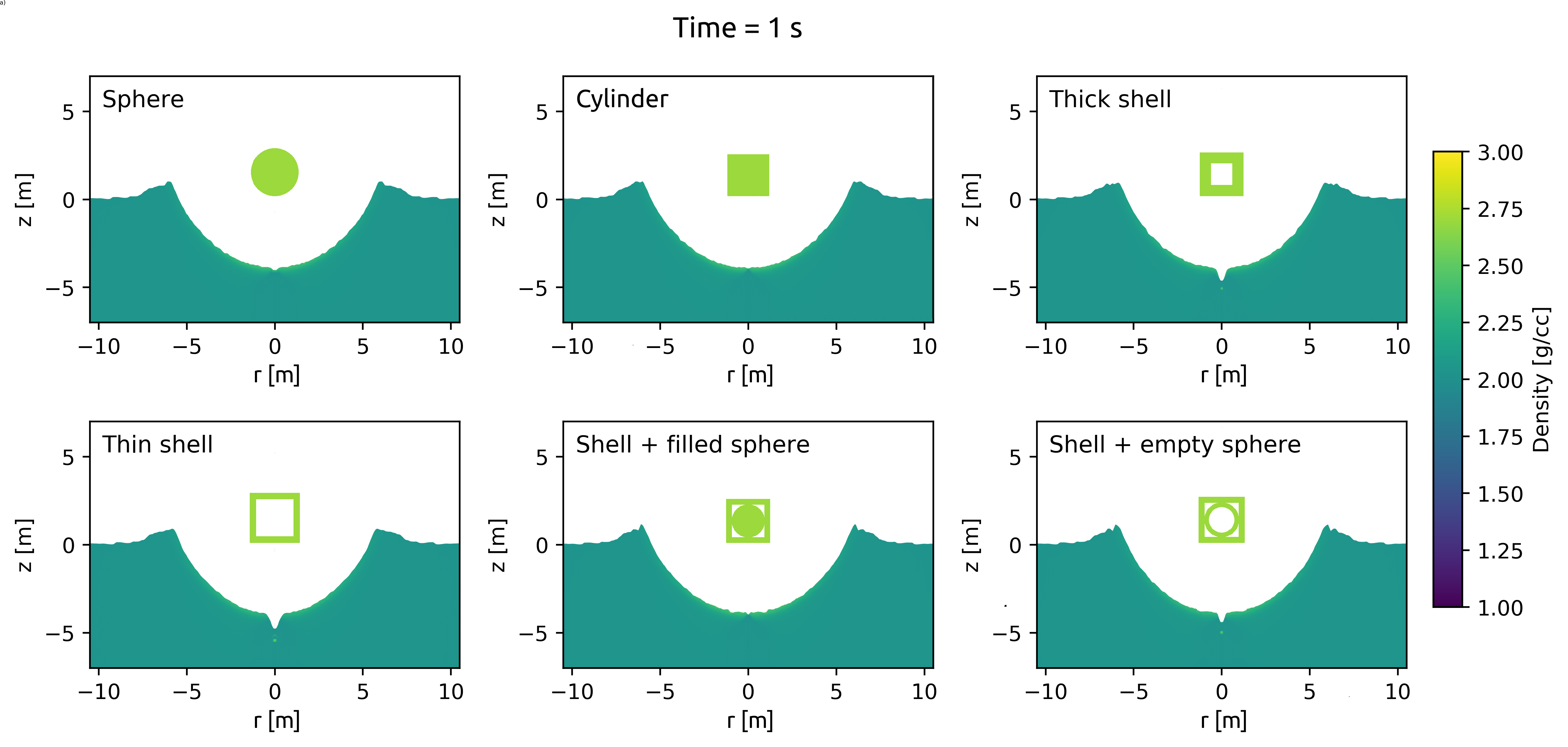}
	\caption{Crater profile at t = 1\,s after the impact. Target cohesion is $Y_{d0}$ = 0.1\,MPa.}
	\label{fig:crater_profile}
\end{figure}

Figure~\ref{fig:crater_profile} shows the crater profiles from impacts of different projectile structures into a $Y_{d0}$ = 0.1\,MPa target, as recorded at 1\,s after the impact. In all cases, the impact produces a bowl-shaped crater, with no significant effects of the projectile geometry. 

\subsection{Ejecta mass-velocity distribution}
\begin{figure}[!h]
	\centering
	\includegraphics[width=\linewidth]{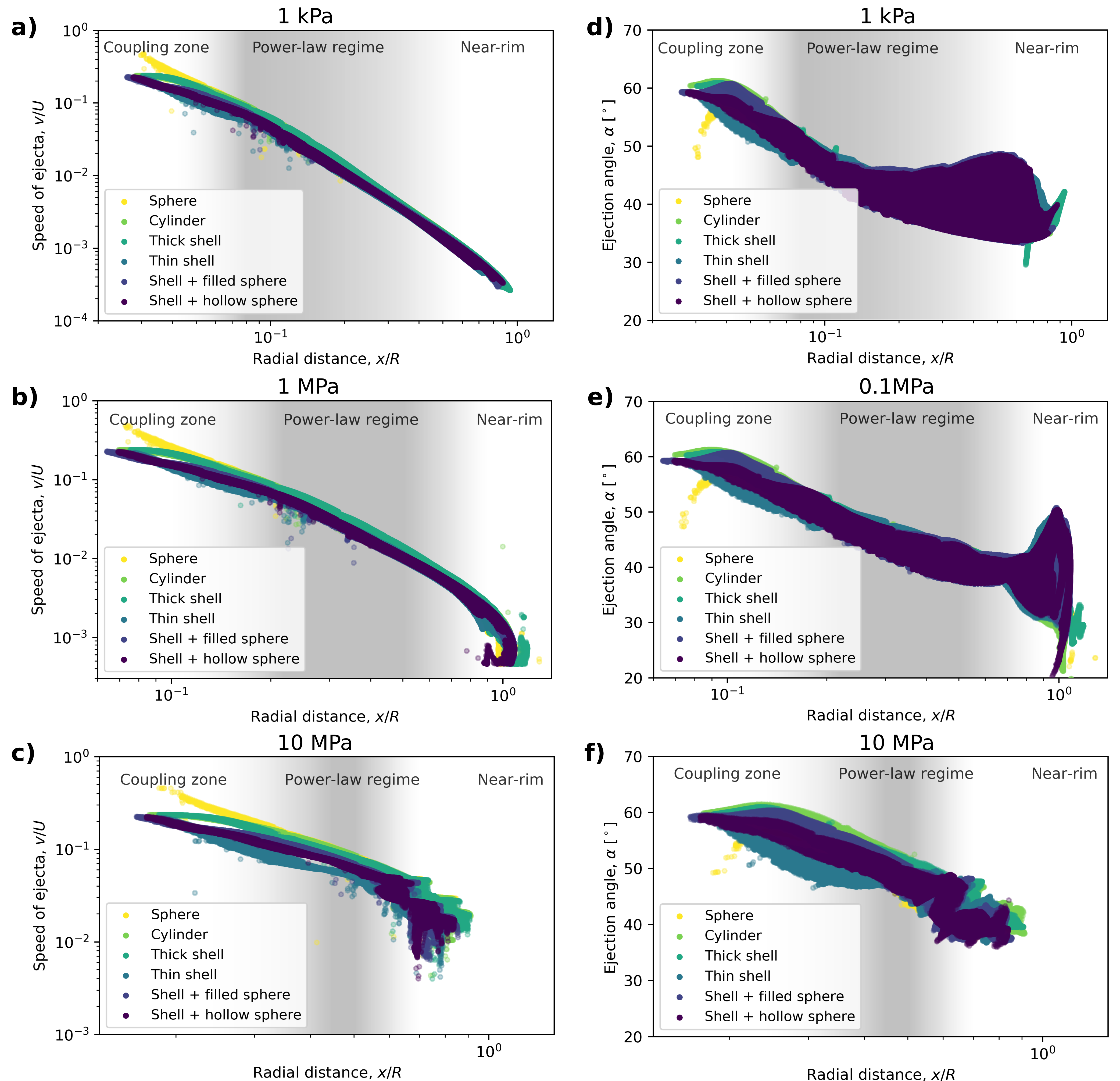}
	\caption{a, b, c) Normalised speed of ejecta and d, e, f) ejection angle as a function of radial distance for different impactor structures. The shaded areas represent the different stages of the ejecta distribution: coupling zone, power-law regime and rear-rim (see \cite{Housen2011} and \cite{Raducan2019} for details).}
	\label{fig:ejecta}
\end{figure}

Figure~\ref{fig:ejecta}a shows the ejecta speed normalised by the impact velocity, $v/U$ as a function of radial distance, $x$, normalised by the final crater radius (measured at pre-impact level), $R$, produced by impacts with six different shaped projectiles, into a $Y_{d0}$ = 1 kPa target. For a given target strength, the ejecta distributions from the six impact scenarios follows very similar trends, with only minor differences (Fig.~\ref{fig:ejecta}a). Most notable differences are seen for the fast ejecta, in the `coupling zone' \citep{Housen2011}, while `power-law' and `near-rim' regimes, the ejecta follows very similar mass-velocity distributions, regardless of the projectile geometry. In the case of the spherical projectile, the first ejecta to leave the crater displays larger ejection velocities compared to ejecta from the other five impact scenarios. 
A very similar trend in the ejecta speed-radial distance distribution can be seen in the case of the $Y_{d0}$ = 1 MPa (Fig.~\ref{fig:ejecta}b) and the $Y_{d0}$ = 10\,MPa target (Fig.~\ref{fig:ejecta}c).
While there is an interplay between projectile geometry effects and target property effects, projectile effects are more evident in impacts with low cratering efficiencies, where the final crater size is not much larger than the projectile size. Due to the higher target cohesion, the 10\,MPa and the 1 MPa target scenarios have a smaller cratering efficiency compared to the 1\,kPa target, which in turns results in a smaller relative duration of the power-law regime. However, the duration of the 'coupling-zone' ejecta is comparable in all target scenarios. 
Considering that the involved pressures are much higher than the target strength, the target cohesion is not expected to affect the early time ejecta behaviour. However, the cohesion is expected to influence the late-time ejecta behaviour and the total amount of ejected mass.

Figures~\ref{fig:ejecta}d, e and f show the ejection angle as a function of normalised radial distance, for (d) $Y_{d0}$ = 1\,kPa, (e) $Y_{d0}$ = 0.1\,MPa and (f) $Y_{d0}$ = 10\,MPa targets. In the case of the spherical projectile, the material ejected close to the impact point has lower ejection angles compared to the other impact scenarios. In this case, the first material to leave the crater is ejected at angles of $\approx 45^\circ$, then the ejection angle increases to $\approx 60^\circ$. In all the other five impact scenarios the material ejected close to the impact point has an ejection angle of about $\approx 60^\circ$. This difference in the ejecta velocity and ejection angle for the spherical impactor scenario compared to the other projectile geometry scenarios is most likely caused by jetting \citep{Johnson2014}. 

In all impact scenarios, the ejection angle slowly decreases from about $\approx 60^\circ$ at 0.1 $x/R$ to $\approx 40^\circ$ close to the crater rim. Similar ejection angles have been noted by \cite{Luther2018} for spherical projectiles. The total mass of ejected material, normalised by the projectile mass, $M_{ej}/m$, was $\approx$~15 for the 10 MPa target, $\approx$~300 for the 0.1 MPa target, $\approx$ 1100 for the 1 kPa target, and it was not influenced by the projectile shape. The simulation results presented here suggest that the impactor geometry only has an effect on the ejecta close to the impact point, in the `coupling zone' of the ejecta distribution (see \cite{Housen2011} for details). As shown in \cite{Raducan2019}, the ejected momentum originating from the coupling zone represents an increasingly larger proportion of the total ejecta momentum, with increasing cohesion. 

\subsection{Momentum enhancement}
\begin{figure}[!h]
	\centering
	\includegraphics[width=0.33\linewidth]{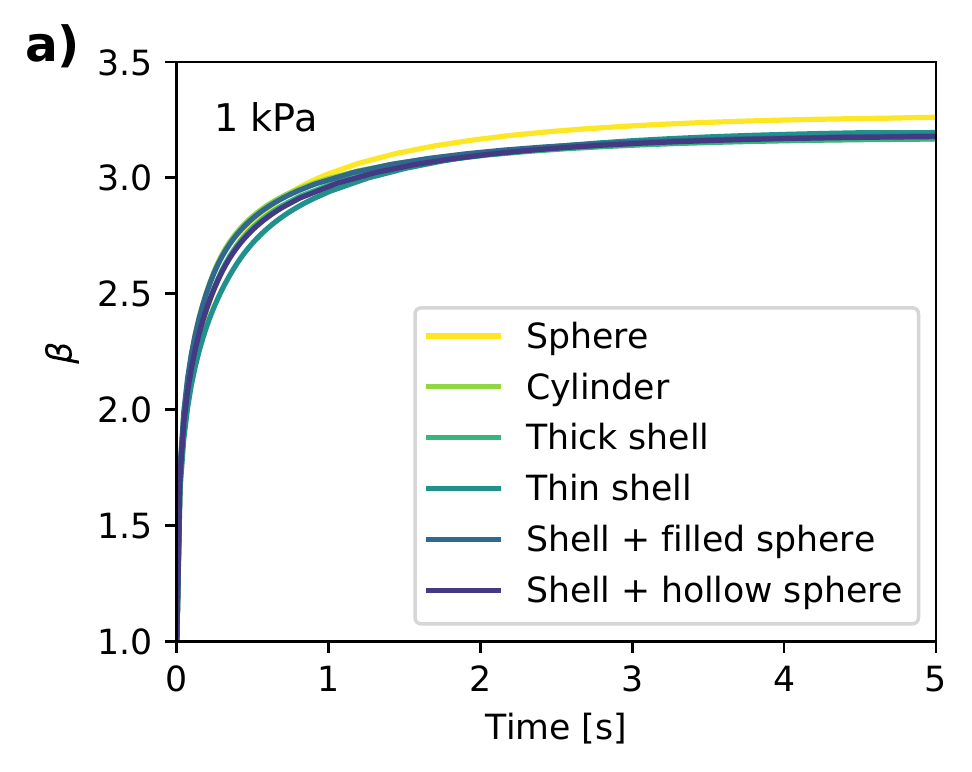}
	\includegraphics[width=0.33\linewidth]{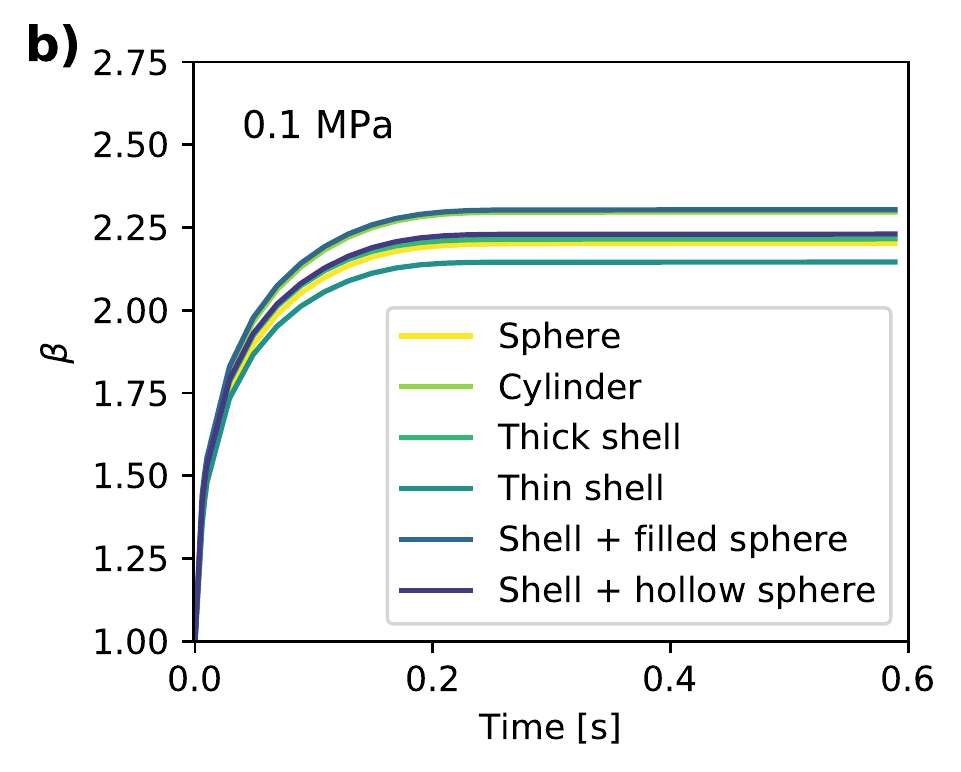}\includegraphics[width=0.33\linewidth]{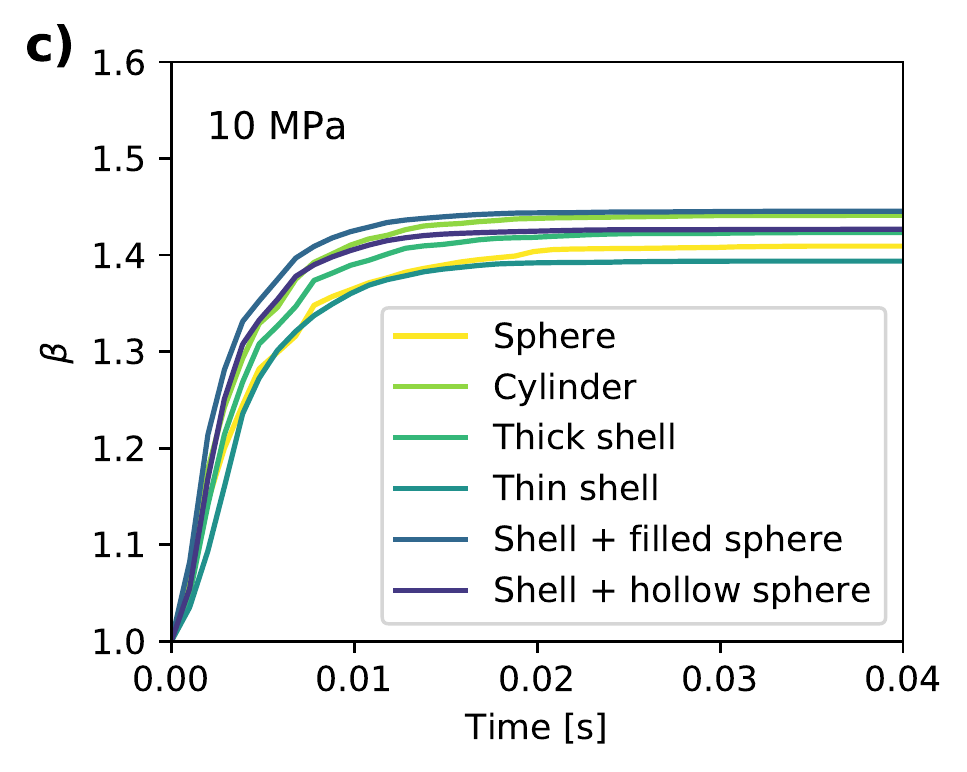}
	\caption{Momentum enhancement, $\beta$, from six different impactor shapes, impacting a basalt target with (a) $Y_{d0}$ = 1 kPa, (b) $Y_{d0}$ = 0.1 MPa, (c) $Y_{d0}$ = 10 MPa.}
	\label{fig:beta}
\end{figure}

An important metric for the DART mission is the amount of momentum transferred to the target by the impact. Figure~\ref{fig:beta} shows the momentum enhancement, $\beta$, as a function of time, from projectiles with different geometries impacting (a) $Y_{d0}$ = 1\,kPa, (b) $Y_{d0}$ = 0.1\,MPa and (c) $Y_{d0}$ = 10\,MPa targets. In the 1\,kPa target scenario, the spherical projectile produces a $\beta$ value that is about 2\% larger than in the other projectile scenarios. The total spread in $\beta$ value for this target scenario was about 3\%. On the other hand, in the case of the 0.1\,MPa and 1\,MPa target scenarios, the sphere, thick shell and shell + hollow sphere projectiles produced very similar $\beta$ values, while the cylinder and the shell + filled sphere projectiles produced an amplification in $\beta$ of about 3\%, compared to the spherical projectile case. The thin shell impactor was less efficient by 2 to 3\% compared to the spherical projectile, in all three target scenarios. 
For the 1\,kPa target scenario, the impactor geometry causes a spread in $\beta$ values of $\approx 3\%$, while for the 0.1\,MPa target scenario, the spread in $\beta$ was about 7\% and in the 1\,MPa target scenario, the spread in $\beta$ was about 4\%. In terms of $\beta-1$, which is the momentum carried away by the ejecta, the spread in values was between 5\% (1\,kPa target) and 15\% (10\,MPa target).  
In summary, the 2D impact simulations show slight variations in the peak pressures experienced by the projectiles and ejection angle distributions, for the simplified projectile geometries studied here, however their effect on the momentum enhancement factor is minimal.


\section{Results from iSALE-3D simulations}

The second study considered vertical impacts with two different projectiles, a sphere and a horizontal rod (aligned in the $x$-direction), modelled in three dimensions. The two scenarios represent simplifications of the spacecraft bus (spherical projectile) and spacecraft bus with solar panels (rod projectile).

\subsection{Crater size and morphology}

\begin{figure}[!h]
	\centering
	\includegraphics[width=0.8\linewidth]{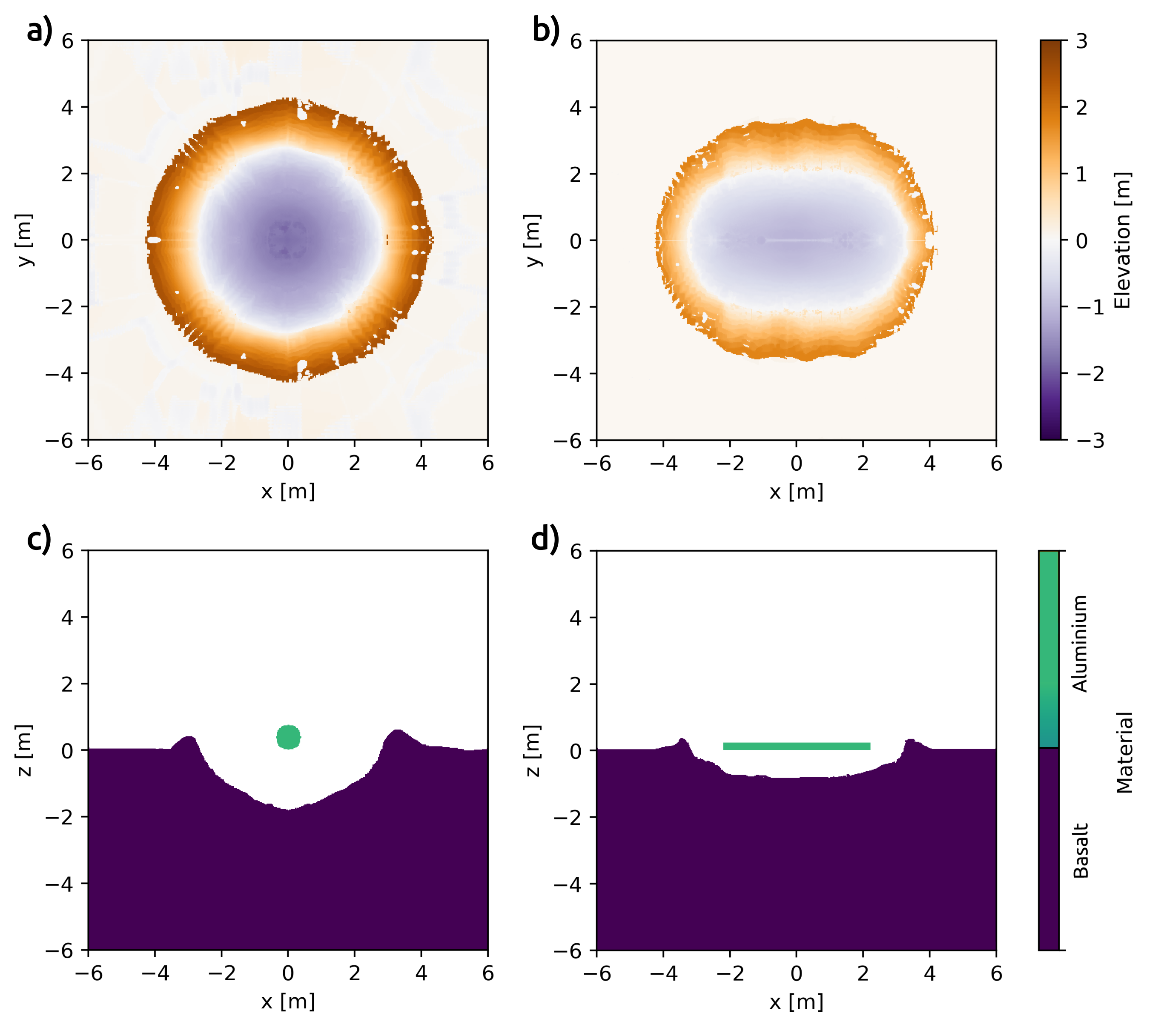}
	\caption{(a, b) Plan view of crater and ejecta curtain evolution at the time when the transient crater is reached ($T$ = 1\,s) and (c, d) crater profiles of the transient craters from iSALE-3D simulations of a sphere and a rod projectile impacting a porous, $Y_{d0}$ = 1 MPa target.}
	\label{fig:profile_3d}
\end{figure}

Figure~\ref{fig:profile_3d} shows plan view of crater and ejecta curtain evolution (a, b) and the crater profile (c, d) from a solid aluminium sphere and solid aluminium rod impacting a $Y_{d0}$ = 1 MPa basalt target. The spherical projectile produces a bowl-shaped spherical crater, with a depth-to-diameter ratio of $d/D \approx$ 0.33. The rod projectile produces an elliptical crater, that is wider in the $x$-direction (the direction parallel to the long axis of the rod) but narrower in the $y$-direction and shallower than the spherical projectile scenario. In this case, the depth-to-diameter ratio is $d/D \approx$ 0.13 for semi-major axes and $d/D \approx$ 0.20 for the semi-minor axes. 

\begin{figure}[!h]
	\centering
	\includegraphics[width=0.49\linewidth]{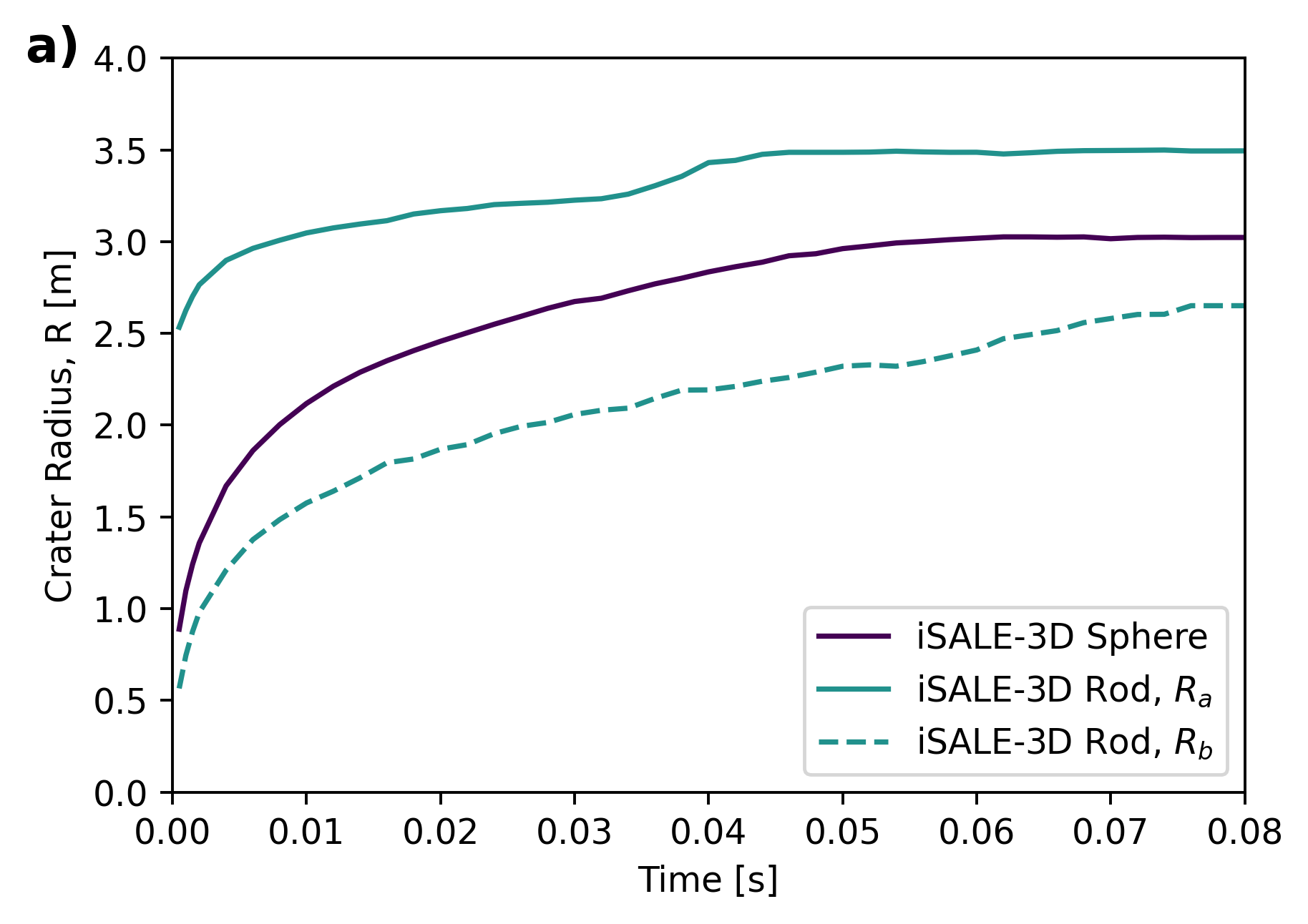}\includegraphics[width=0.49\linewidth]{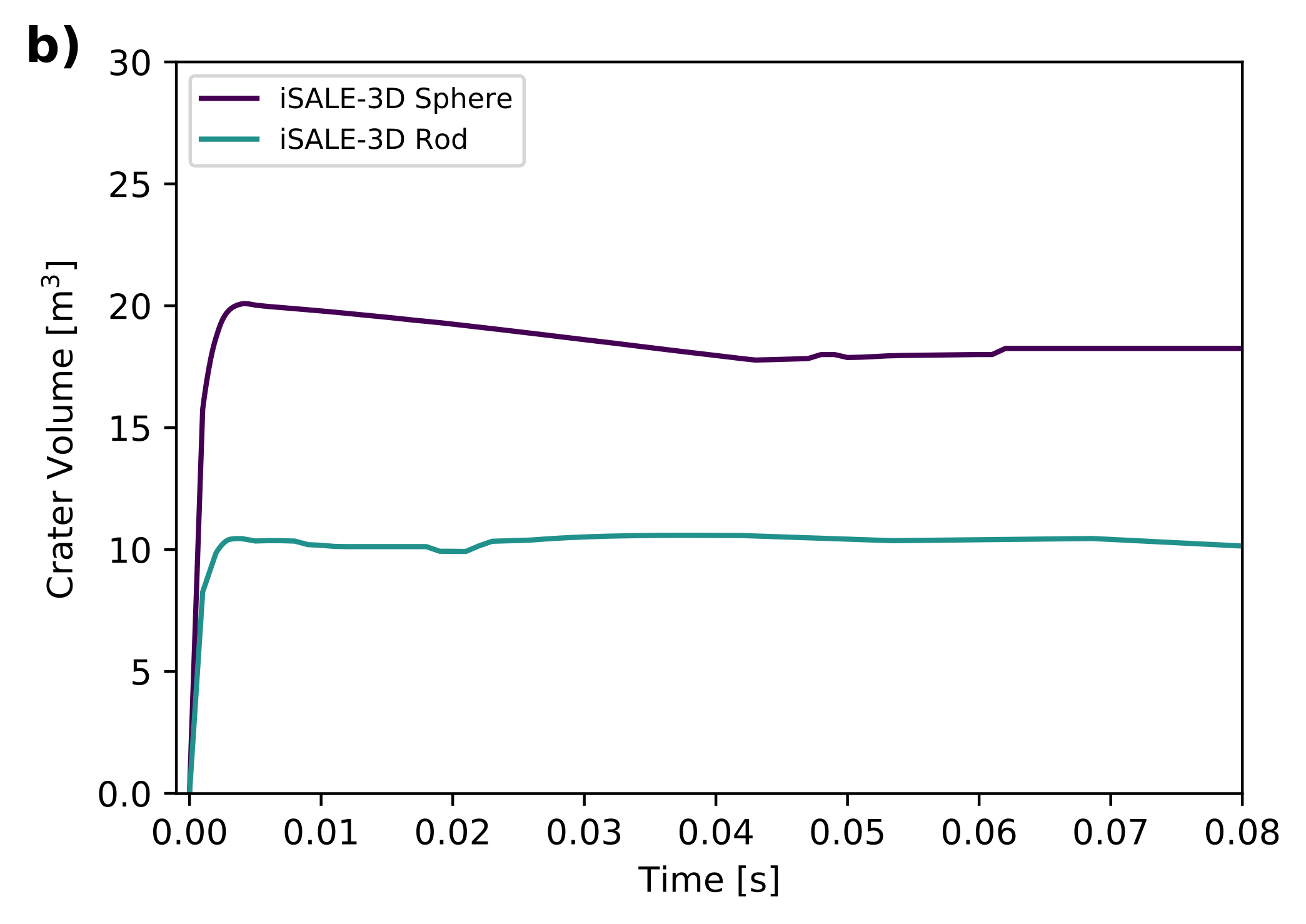}
	\caption{Crater growth from iSALE-3D simulations of a sphere and a rod projectile. a) Crater radius. The spherical projectile produces a circular crater of radius $R$, while the Rod projectile produces an elliptical crater of semi-major axis $R_a$ and semi-minor axes $R_b$. b) Crater volume.}
	\label{fig:cratergrowth3d}
\end{figure}

Figure~\ref{fig:cratergrowth3d} shows the growth of the crater (a) radius and (b) volume, as a function of time, for the two impact scenarios. The spherical projectile produces a circular crater with a radius of $R \approx$~3\,m and a volume of $\approx$~20\,$m^3$. The rod projectile produces an elliptical crater with a semi-major axis, $R_a \approx$ 3.5\,m and a semi-minor axis, $R_b \approx$~2.7\,m. The volume of the elliptical crater was $\approx$~10\,$m^3$, twice as small as the spherical projectile crater. 

\subsection{Ejecta mass-velocity distribution}
\begin{figure}[!h]
	\centering
	\includegraphics[width=\linewidth]{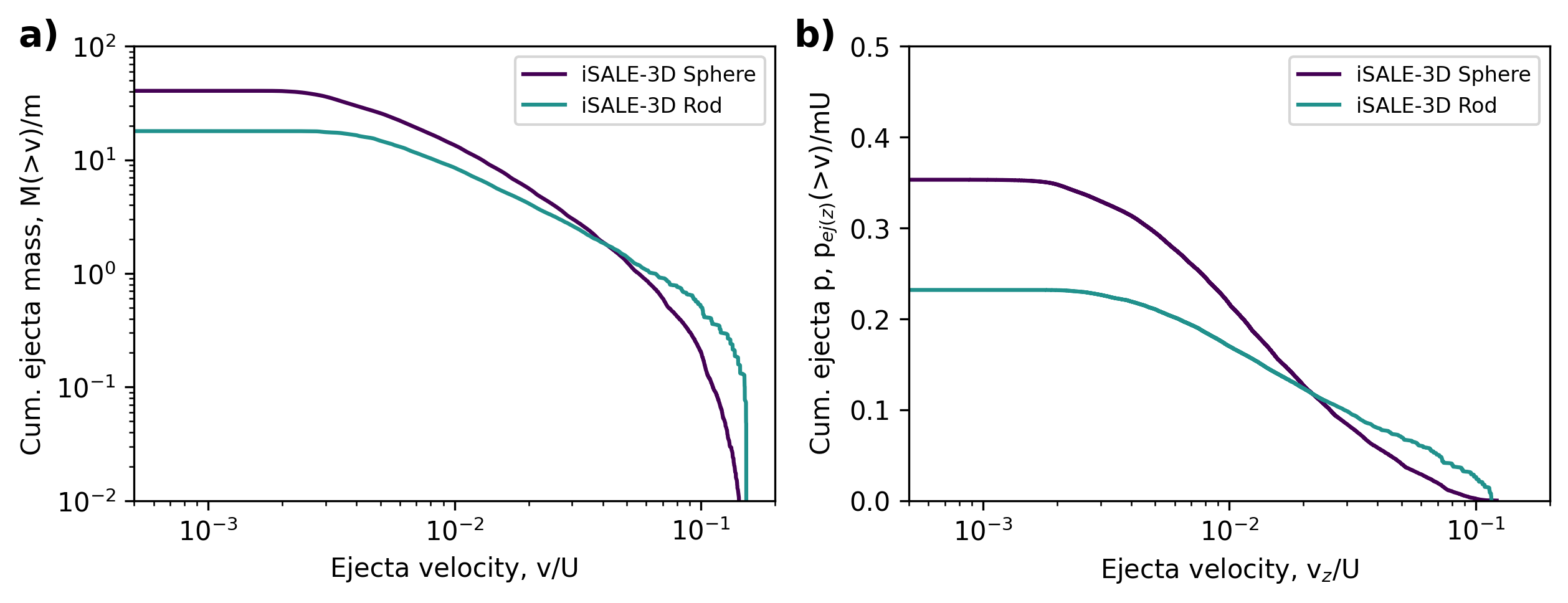}
	\caption{a) Cumulative mass of ejected particles at speeds grater than $v$, as a function of normalised ejection speed $v/U$, for two projectile shapes: sphere and rod, modelled in three-dimensions. b) Cumulative vertical ejecta momentum, normalised by the momentum of the projectile, $mU$, as a function of vertical ejection velocity, normalised by the impactor velocity, $v_z/U$. The value where the ejecta distribution intersects the $y-$ axis represents the total momentum carried away by the ejecta, that contributes to the impact momentum transfer, $\beta-1$.}
	\label{fig:ejecta_3d}
\end{figure}

Figure~\ref{fig:ejecta_3d}a shows the mass-velocity distribution from iSALE-3D simulations of a solid aluminium sphere and of a long aluminium rod impacting a porous, 1 MPa, basalt target. The rod projectile ejects more mass at high velocities ($v/U>0.1$) compared to the spherical projectile. On the other hand, the spherical projectile has a larger efficiency and there is more mass being ejected overall, thus at lower ejection velocities. 

Figure~\ref{fig:ejecta_3d}b shows the normalised cumulative ejecta momentum as a function of vertical ejection velocity. In the case of the rod impactor, the larger amount of ejected mass at high velocities (compared to the spherical projectile scenario) translates to a larger amount  momentum being imparted by the ejected particles at high velocities. However, the spherical projectile produced an overall cumulative ejecta momentum that is about 10\% larger than in the rod impactor scenario. 

\section{Discussion}

Here we first modelled simple projectile structures with axial symmetry, impacting targets with three different cohesion configurations, here referred to based on the damaged cohesion, $Y_{d0}$ = 1 kPa, 0.1 MPa and 10 MPa. For all three target scenarios, the projectile geometry has minimal effects on the crater radius and crater volume (less than 5\%). These results suggest that projectiles with similar footprint surface areas do not influence the crater size significantly, even in the case of strong targets, where the crater efficiency is much lower. 


When projectiles with different footprint surface areas were considered, i.e., spherical and rod projectiles modelled in 3D, the differences in crater dimensions were clearer. 
However, our results suggest that for projectile geometries with similar surface areas at the point of impact there are subtle differences in the ejecta speed-radial distance distribution and in the angle of the ejecta generated from these impacts. These differences cause subtle amplifications and reductions in the momentum enhancement, compared to a spherical projectile scenario. Here we did we did not see a clear trend in the change in $\beta$ values with projectile geometry, and the spread in $\beta$ values was less than 5\%.

The projectile effects are most evident in the fastest ejecta, produced at very early times close to the contact zone when target strength effects are negligible. Hence, these effects are expected to be independent of target strength. As $\beta$ is dominated by the slowest ejecta, projectile effects on deflection efficiency are expected to diminish with decreasing target strength.


This behaviour of the fast ejecta caused by the projectile geometry might be visible in the early crater ejecta plume. Information about the DART crater ejecta plume will be available and will be provided by LICIACube \citep{Cheng2020, Dotto2021}. LICIA (Light Italian Cubesat for Imaging of Asteroids) is the Italian Space Agency (ASI) contribution to the DART mission. The CubeSat will be carried by the DART spacecraft and will be released in the vicinity of the Didymos system before the impact. The main aim of the CubeSat is to take images of the ejecta plume, over a range of angle phases, at 136--163\,s after the impact. 

The timescale of the `coupling zone' ejecta produced in our simulations is too short and would be too transparent to be observed at $\approx$ 150\,s after the impact. However, the transition between the `coupling zone' and the power-law regime is gradual and the duration of each stage is dependent on the projectile and target properties. Previous experimental and numerical studies \citep[e.g.,][]{Schultz1988, Hermalyn2010, Raducan2019} of vertical impacts showed that while a very small proportion of the crater growth occurs in the `coupling zone', its duration increases relative to the total crater formation time as the impactor increases in size and velocity. Here we can also make an analogy with impact cratering experiments with projectiles with different densities \citep{Hermalyn2011}: dense projectiles (or projectiles with less voids) penetrate deeper into the target and the coupling of the energy and momentum occurs later in the cratering process, while an under-dense projectile (a projectile with more voids) remains closer to the surface and the coupling occurs quicker.

Moreover, previous numerical simulations \citep[i.e.,][]{Raducan2019} showed that target porosity (or more precisely the ratio of the projectile and target porosities) also affects the size of the `coupling zone'. However, target strength parameters such as the cohesion and the coefficient of internal friction do not. While a weaker target translates to longer transient crater growth times (i.e., ranging from $\approx$ few minutes for a 50\,Pa target, up to $\approx$ 2\,h for a cohesionless target \citep{Raducan2021b}), nonetheless it is important to note that this does not translate to a longer `coupling time'. Therefore, the strength of the target is not expected to influence the projectile geometry's effect on the ejecta cone.

However, due to the projectile to target density ratio, which is not yet known, differences might still be visible at the times of the LICIACube images the ejecta plume. In other words, the observed ejecta plume produced by a kinetic impactor with a complex shape (e.g., a spacecraft) may appear different to what would be expected for ejecta produced by a homogeneous spherical projectile. A complex projectile is expected to exhibit crater ejecta with lower ejection speeds and higher ejection angles compared to a simple sphere. 

The results presented here are only the first step towards understanding the effects of projectile geometry on crater morphology and momentum enhancement from a kinetic impactor. In the case of the DART spacecraft, the projectile geometry (i.e., the spacecraft geometry) is much more complex, and full 3D models of the entire spacecraft, including solar panels, need to be considered. In this aspect, several works are in progress (e.g., \cite{Stickle2018}), however such simulations are very computationally demanding and are typically restricted to very specific scenarios (e.g., only to impacts into targets with low cratering efficiency, such as very strong targets).

\section{Conclusions}

The DART mission will impact Didymos's secondary, Dimorphos, at the end of 2022 and cause a change in the orbital period of the Dimorphos around the primary. For simplicity, previous studies aimed at quantifying the consequences of the DART impact have used a spherical projectile. However, the geometry of a kinetic impactor is more complex. Here we conducted a systematic study to determine the effects of simple projectile geometries to the crater size and morphology and to the ejecta mass-velocity distribution.

We found that for geometries with similar footprint surface area, the projectile geometry affects crater radius and crater volume by less than 5\%. Moreover, the projectile geometry can both amplify and reduce $\beta$, but by less than 5\% compared to a spherical projectile. The more prominent differences produced by the different projectiles are seen in the ejected mass-radial distance and ejection angle distribution of the fast ejecta. LICIACube, which will image the DART impact ejecta plume, $\approx$~150\,s after the impact, might record crater ejecta that has higher ejection angles compared to what is expected from a spherical projectile. However, the exact differences will depend on the target porosity and on the imaging conditions.

We also considered the difference between a sphere and a horizontal rod impactor geometry (here considered a simplification of a spacecraft with solar panels) that used a full 3D impact simulation. The rod projectile, which had a very different footprint surface area and aspect ratio to the spherical impactor, produced an elliptical crater that was about 50\% shallower than the crater produced by the spherical impactor. The momentum enhancement from the rod impact scenario was also $\approx$~10\% lower than in the spherical projectile scenario. 

Our results suggest that for simple projectile geometries, differences in projectile geometry have only a very small effect on the momentum transfer from a kinetic impactor, however further studies, with more realistic projectile geometries (i.e., a spacecraft) are needed to study this in more detail. 

\section{Acknowledgements}
We gratefully acknowledge the developers of iSALE (www.isale-code.de), including Kai W\"unnemann, Dirk Elbeshausen, Boris Ivanov and Jay Melosh. This work has received funding from the European Union’s Horizon 2020 research and innovation programme, grant agreement No. 870377 (project NEO-MAPP) and the UK's Science and Technology Facilities Council (STFC) (Grant ST/N000803/1). J. M. Owen's work was performed under the auspices of the U.S. Department of Energy by Lawrence Livermore National Laboratory under Contract DE-AC52-07NA27344; LLNL-JRNL-824958.

\section{Data availability}
Additional supporting information (tables, model outcomes) will be archived on GitHub and provided at the time of the publication as a DOI.

\newpage
\bibliographystyle{apa}  
\bibliography{refs} 

\end{document}